\documentclass{article}
\usepackage{ijcai21}

\usepackage{times}
\usepackage{amsmath}
\usepackage{amssymb}
\usepackage{soul}
\usepackage{url}
\usepackage[hidelinks]{hyperref}
\usepackage[utf8]{inputenc}
\usepackage[small]{caption}
\usepackage{graphicx}
\usepackage{amsthm}
\usepackage{amsmath}

\usepackage{booktabs}
\usepackage{algorithm}
\usepackage{algorithmic}
\usepackage{acronym}
\acrodef{O/O}{owner/operator}
\acrodefplural{O/O}[O/Os]{owners/operators}
\acrodef{SSN}{Space Surveillance Network}
\acrodef{CDM}{conjunction data message}
\acrodef{TCA}{time of closest approach}
\acrodef{PP}{Poisson Process}

\newtheorem{example}{Example}
\newtheorem{theorem}{Theorem}
\newtheorem{definition}{Definition}

\title{Conjunction Data Messages behave as a Poisson Process}

\author{
Francisco Caldas$^{1,2}$
\and
Cl\'{a}udia Soares$^3$\and
Cl\'{a}udia Nunes$^{2}$\and
Marta Guimar\~{a}es$^1 $\and
Mariana Filipe$^1$ \and
Rodrigo Ventura$^2$
\affiliations
$^1$Neuraspace\\
$^2$ Instituto Superior T\'{e}cnico and CEMAT\\
$^3$ NOVA School of Science and Technology
\emails
\{francisco.caldas,marta.guimaraes,mariana.filipe\}@neuraspace.com,
claudia.soares@fct.unl.pt, cnunes@math.tecnico.ulisboa.pt, rodrigo.ventura@isr.tecnico.ulisboa.pt
}

\begin{document}

\maketitle

\begin{abstract}
    Space debris is a major problem in space exploration. International bodies continuously monitor a large database of orbiting objects and emit warnings in the form of conjunction data messages. An important question for satellite operators is to estimate when fresh information will arrive so that they can react timely but sparingly with satellite maneuvers. We propose a statistical learning model of the message arrival process, allowing us to answer two important questions: (1) Will there be any new message in the next specified time interval? (2) When exactly and with what uncertainty will the next message arrive? The average prediction error for question (2) of our Bayesian Poisson process model is smaller than the baseline in more than 4 hours in a test set of 50k close encounter events. 
\end{abstract}

\section{Introduction}

Since the early 1960s, the space debris population has extensively increased~\cite{radtke2017interactions}. It is estimated that more than 34,000 objects larger than 10 centimetres, and millions of smaller pieces, exist in Earth's orbit \cite{esa_2021}. Collisions with debris give rise to more debris, leading to more collisions in a chain reaction known as Kessler syndrome~\cite{krag20171}. To avoid catastrophic failures, satellite \acp{O/O} need to be aware of the collision risk of their assets~\cite{le2018space}. Currently, this monitoring process is done via the global \ac{SSN}. To assess possible collisions, a physics simulator uses \ac{SSN} observations to propagate the evolution of the state of the objects over time~\cite{horstmann2017investigation,mashiku2019recommended}. Each satellite (also referred to as target) is screened against all the objects of the catalogue in order to detect a conjunction, i.e., a close approach. Whenever a conjunction is detected between the target and the other object (usually called chaser), \ac{SSN} propagated states become accessible and a \ac{CDM} is issued, containing information about the event, such as the \ac{TCA} and the probability of collision. Until the \ac{TCA}, more \acp{CDM} are issued with updated and better information about the conjunction. Roughly in the interval between two and one day prior to \ac{TCA}, the \acp{O/O} must decide whether to perform a collision avoidance manoeuvre, with the available information. Therefore, the \ac{CDM} issued at least two days prior to \ac{TCA} is the only guaranteed information that the \acp{O/O} have and, until new information arrives, the best knowledge available.

Several approaches have been explored to predict the collision risk at \ac{TCA}, using statistics and machine learning~\cite{acciarini2021kessler}, but only a few were developed with the aim of predicting when the next \ac{CDM} is going to be issued. Very recently, \cite{pinto-2020-automate} developed a recurrent neural network architecture to model all \ac{CDM} features, including the time of arrival of future \acp{CDM}. GMV is currently developing an autonomous collision avoidance system that decides if the current information is enough for the \acp{O/O} to decide, or if they should wait for another CDM to have more information.
However, the techniques used are not publicly available. In this work, we propose a novel statistical learning solution for the problem of modeling and predicting the arrival of \acp{CDM} based on a homogeneous \ac{PP} model, with Bayesian estimation. We note that standard machine learning and statistical learning solutions are data-hungry and cannot model our problem, suffering from a special type of data scarcity. Further, as our application is high-stakes, we require confidence or credibility information added to a point estimate. 

\paragraph*{Background.} The present work formalizes the problem of modelling and predicting the arrival of a new \ac{CDM} with a probabilistic generative model. The present formulation has the advantage of providing a full description of the problem, stating clearly the required assumptions. Our proposed model shows high accuracy, decreasing error in more than 4 hours, when compared with the baseline, for predicting the next event.
\begin{figure}[t]
    \centering
    \includegraphics[scale=0.6]{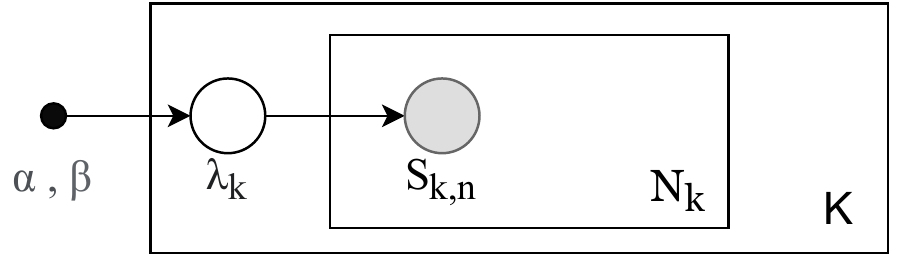}
    \caption{Graphical model representation of the proposed model. The boxes are plates representing replication of the structure. The outer plate represents the dataset of $K$ events, and the inner one the repeated prediction of $N_k$ occurrences in each event.}
    \label{fig:graph}
\end{figure}
We leverage on a data model depicted in Figure~\ref{fig:graph}. Probabilistic Graphical Models~\cite{koller2009probabilistic,bishop2006pattern} is a probabilistic framework that allows for highly expressive models, while keeping computational complexity to a minimum. A few statistical learning developments like the Latent Dirichlet Allocation for topic modelling~\cite{blei2003latent}, are now the basis for deep learning-powered probabilistic models like the Variational Autoencoder~\cite{kingma2019introduction} and Normalizing flows~\cite{kobyzev2020normalizing}, and as a means of creating modular and interpretable machine learning models~\cite{murdoch2019definitions}. We leverage on the homogeneous Poisson Point Process~\cite{kulkarni2016modeling}, a counting process of messages occurring over continuous time. Thus, we assume that the inter-CDM times are independent and identically distributed random variables with exponential distribution with a rate~$\lambda$. The homogeneity property comes from an event rate constant in time. Phenomena like radioactive decay of atoms and website views are well modeled by this process~\cite{ross2014introduction}. The homogeneity is, nevertheless, a limitation that will be addressed in future work, as there might be some temporal trend in the data. Not considering it, though, delivers a fast to compute solution that can outperform a standard baseline.

\section{Problem Formulation}

Our generative probabilistic model of the CDM arrival process will allow answering some questions as (1) what is the probability that, during the decision phase, the roughly one day that the operator has to decide to perform an avoidance manoeuvre, there are going to be received more CDMs with up-to-date information, or (2) what is the best estimate for the next CDM arrival time, and the uncertainty of this prediction.
\begin{definition}
Following the literature, we define the stochastic quantities, with $k$ as the id of the object:
\begin{itemize}
    \item $X_{n+1}^k$ time between CDM $n$ and CDM $n+1$, where $X_{n+1}^k \sim Exp(\lambda_k)$, i.e., $X_{n+1}$ has an exponential distribution of rate $\lambda_k$;
    \item $N^k(t) $ = number of CDMs received in the interval (0,t], where  $N^k(t) \sim \text{Poisson}(\lambda_k t)$, where $\lambda_k$ is the rate of CDMs issued per day;
    \item $S_n^k  = \sum_{i=1}^n X_i^k$ as time of occurrence of the $n_{th}$ CDM;
    \item as $\{ X_i^k,i=1,..,n\}$ are i.i.d., it follows that $S_n^k \sim Gamma(n,\lambda_k)$.
\end{itemize}
\end{definition}
We are interested in the probability of receiving one or more CDMs in the time interval from the last available observation $s_n^k$, until a constant security threshold $(t_{sec})$. In this application, we consider 1.3 days prior to TCA, and the interval will be $(s_n^k,t_{sec}]$.
\begin{align}
    P(N^k(t_{sec}-s_n^k) \geq 1) &= 1 - P(N^k(t_{sec}-s_n^k) =0)  \label{eq:prob} \\ 
   & = 1-P(\text{Poisson}( \lambda_k(t_{sec}-s_n^k)=0) \nonumber
\end{align}
To determine the most likely time for the next CDM to be issued, for each event, using the fact that inter-CDM times have Exponential distribution:
\begin{align*}
  &S^k_{n+1} = S^k_n + X^k_n \qquad \qquad \qquad 
  X^k_n \sim Exp(\lambda_k)  \\
& P(S^k_{n+1} -S^k_n \leq s| S^k_{n}=s_n^k) =1- e^{-\lambda (S_{n+1}-s^k_{n})}
\end{align*}

With these assumptions, we are left with the need to estimate $\lambda_k$,  for each event $k$. In the next section we explain the approach.

\section{Maximum a Posteriori}

Because of the small number of CDMs in each event, to have a statistically significant estimation of $\lambda_k$, and to use information from the other events, we adopt the Bayesian model in Figure~\ref{fig:graph}. More specifically, we use the extensive amount of events in the dataset to get an informative prior, and then we update that distribution with the specific inter-CDMs times of each event. We end up with a posterior distribution for $\lambda_k$, of which we can extract a Maximum a Posteriori (MAP), $\hat{\lambda}_k$ \cite{ruggeri}.

The posterior $\lambda$ distribution is defined as such:
\begin{enumerate}
    \item $\lambda$ has prior Gamma($\alpha$,$\beta$) where $\alpha$ and $\beta$ are hyperparameters estimated using the events in the train data;
    \item The likelihood function, considering we have a set of $X_i$ for each event is $$l(\lambda_k|X_1^k,X_2^k,...,X_n^k) = \lambda_k^n e^{-\lambda_k \sum^n_{i=1}X_i^k}$$
    \item With $T=\sum^n_{i=1}X_i$, the posterior can therefore be written as \cite{kulkarni2016modeling}:
\begin{align}
    f(\lambda_k| n,T) \sim \text{Gamma}( \alpha + n, \beta + T).
    \label{eq:po}
\end{align}

\end{enumerate}

\begin{figure}[t]
    \centering
    \includegraphics[width=\columnwidth]{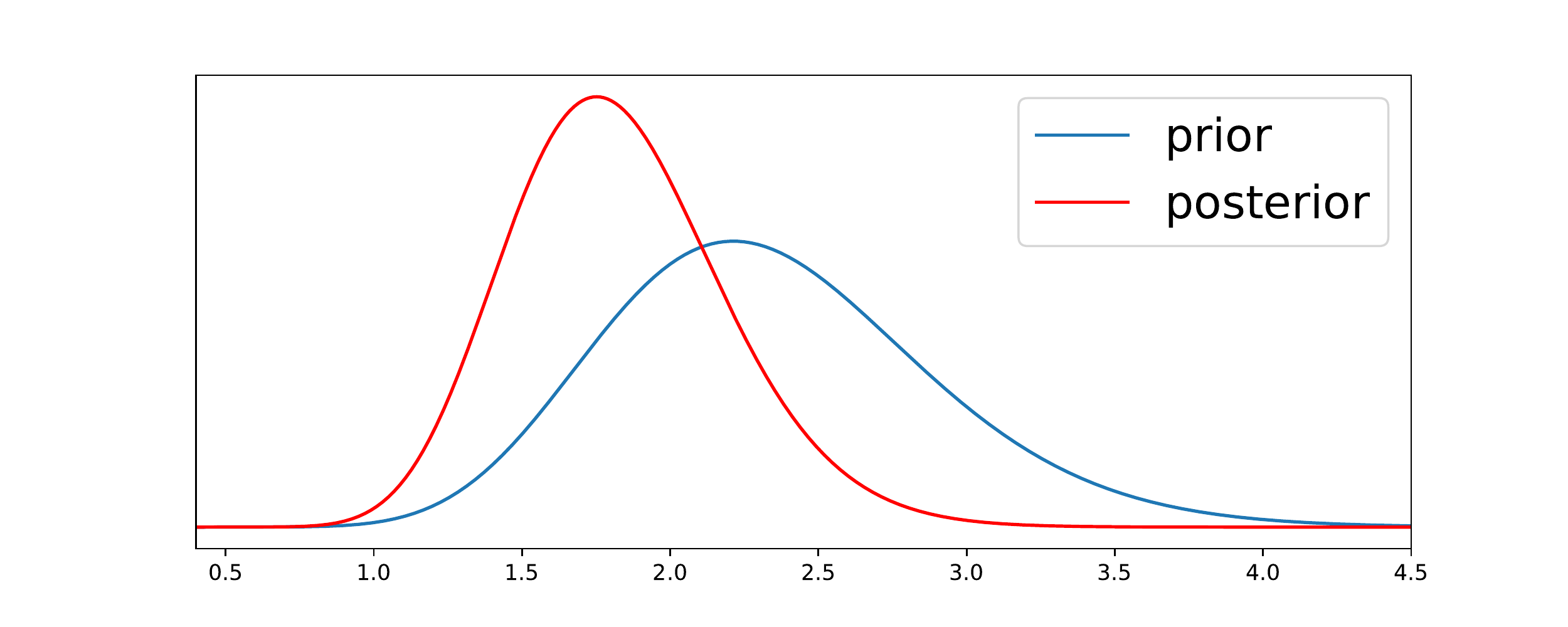}
    \caption{Prior distribution of $\lambda_k$, and a posterior distribution for a single event in the train data. We can see that while informative, the prior has enough variance that with few updates there is a shift in the distribution of the posterior rate of arrival ($\lambda_k$). We can also see that the posterior is narrower, meaning that the credible interval of the posterior is smaller.}
    \label{fig:my_label}
\end{figure}

\section{Results}

The first result we show is the prediction of the time of the next CDM.
\begin{table}[H]
\centering
\begin{tabular}{llll}
\hline
  & MAE & MSE & RMSE \\
\hline
Bayesian PP  &  0.15170   &    0.06282  & 0.25064 \\
Classic PP &  0.17581   &   0.08789  & 0.29646 \\
Baseline & 0.26075 &   0.19105 & 0.43709 \\
\hline
\end{tabular}
\caption{Mean Average Error (MAE) and Mean Squared Error (MSE) for next CDM time prediction in days.}
\label{tab:plain}
\end{table}

\begin{figure}[t]
    \centering
    \includegraphics[width=\columnwidth,]{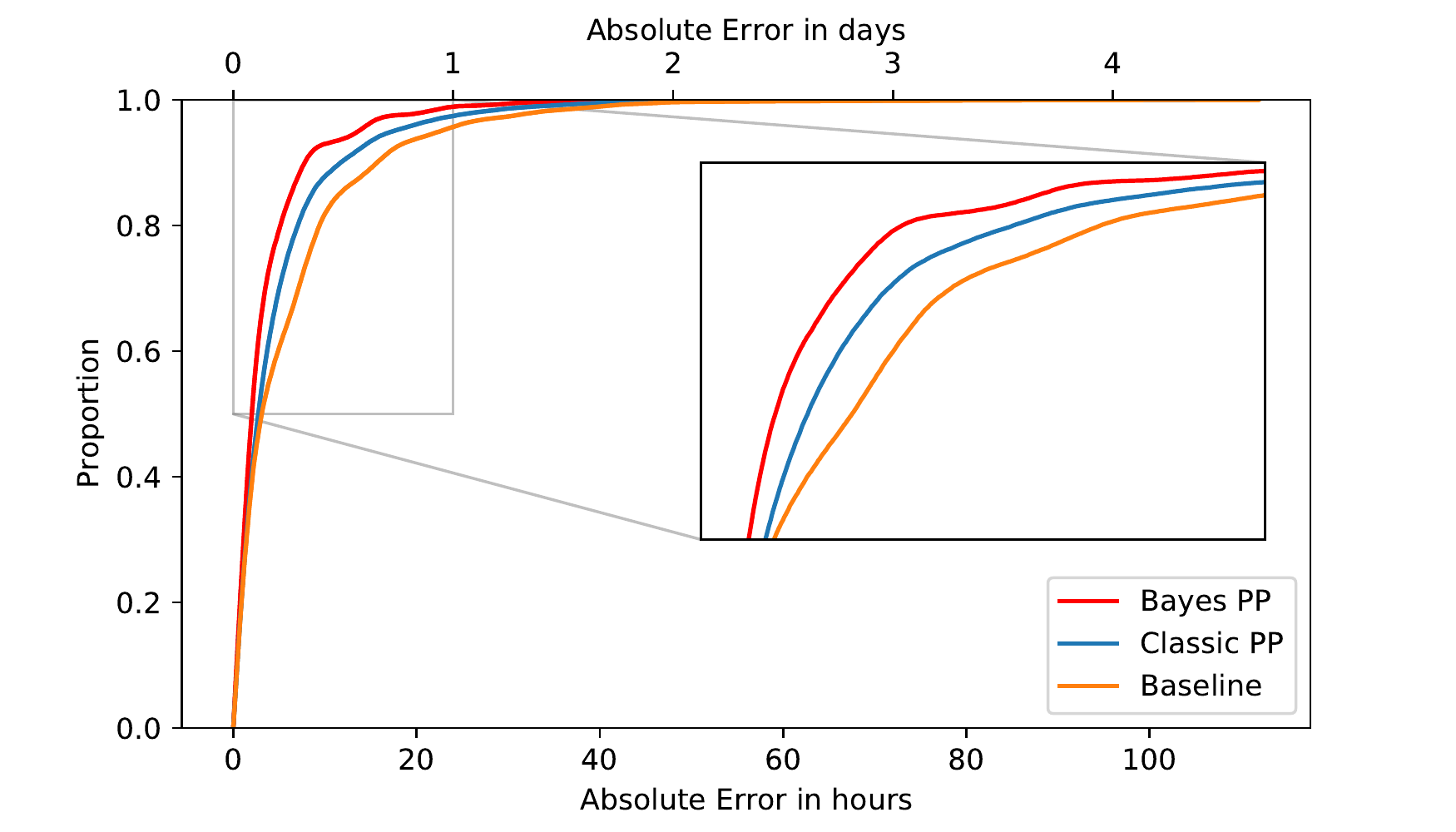}
    \caption{Empirical CDF for each of the models. There is a perceivable difference in the predictive capability of each model. Because the three models use the same formula do predict the expected time of the next CDM, the difference lies in parameter determination, that is, the rate of arrival $\lambda_k$.}
    \label{fig:cdf}
\end{figure}

These results were obtained using an unbiased test data of 50000 independent events, that was not seen during the study and was not used to derive the hyperparameters. The Root Mean Squared Error (RMSE) for the proposed bayesian PP is $0.2506$ days, which corresponds to $6$h.
Comparing the Bayesian PP with the classical estimation of the parameter of the \ac{PP}, we note that we expect better accuracy, as the number of CDMs for each event is small, and thus, the classical approach is more sensitive to extreme values.
As can be seen in Figure~\ref{fig:cdf}, the proposed model outperforms the classic counterpart and the baseline. The baseline is the assumption that the previous inter-CDM time is the best estimator of the next inter-CDM time.

In order to confirm the good performance of our approach, we compare the estimated probability of receiving a CDM in a decision interval with the empirical probability. The estimated probability is the probability obtained using equation \eqref{eq:prob}. To obtain the empirical probability, the events are grouped by the estimated probability intervals as presented in Table \ref{tab:prop}. Then,  by computing ratio of events in each group that actually receive a CDM in that time interval, we obtain an empirical probability.
It is important to note that for each group, the probabilities are not uniformly distributed, however, we can still make some observations and the results are indeed promising. We note that there is a small positive drift between the estimated probability under the Poisson assumption and the empirical one, meaning that the estimated probability is conservative when compared to reality. We also note that, as it should be, a lower estimated probability will indeed represent a lower empirical probability, because along the groups, the empirical probability increases.

\begin{table}[b]
\centering
\begin{tabular}{llr}
\toprule
Estimated Prob.  &  Empirical Prob.  & Deviation \\
\midrule
(0.704, 0.753]    & 0.7647  &  0.0117  \\
(0.753, 0.803]   & 0.83992  &    0.03692     \\
(0.803, 0.852]  & 0.90365 &  0.05165     \\
(0.852, 0.901]  & 0.9401  &  0.0391  \\
(0.901, 0.951]  & 0.96740  & 0.0164  \\
(0.951, 1.0]   & 1.0  & 0\\
\bottomrule
\end{tabular}

\caption{Grouping of events by calculated probability, and the empirical results for each group. The Estimated probability is not over-confident when determining if a CDM is going to be issued during the decision interval.}
\label{tab:prop}
\end{table}

\begin{figure}[t]
    \centering
    \includegraphics[width=0.9\columnwidth]{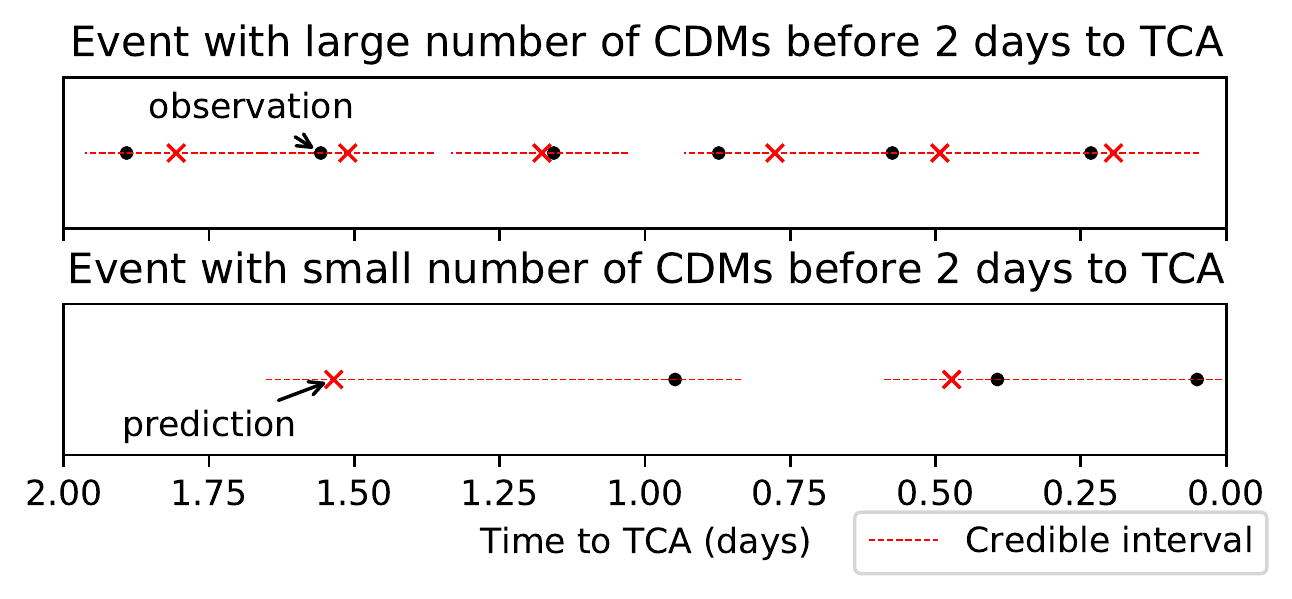}
    \caption{Example of the predictive capability of the model for different examples. Top panel uses 15 observations to update the prior. Bottom panel uses 3 observations. It is of note that for events with fewer observations to update the prior, as is the case of the Bottom event, the credible interval is bigger, to represent a higher uncertainty in the prediction. 90\% credible interval in both cases.}
    \label{fig:examples}
\end{figure}

In Figure~\ref{fig:examples} we can see two examples of the temporal prediction for two events in our test data, with 90\% credible intervals. This figure shows that when the number of observations used for the Bayesian estimation is not loo low (15, top panel), the observed and estimated times are close, and more important from a statistical point of view, the observed value is contained within the credibility intervals.
In the lower panel case, the credibility intervals have a large range (spanning over almost one day), and the observations and predictions are far apart.

\section{Conclusion and Future Work}

In conclusion, we found that this real-life problem in space situational awareness can be successfully modeled by a stochastic process, and that by using a Bayesian scheme, we can overcome data scarcity. By modeling this problem as a \ac{PP}, we can estimate the arrival of CDMs during the decision period, which can be used in practice to aid with expert decision process, helping the operator to confidently delay the manoeuvre decision until new information is received.

Our results have shown that our dataset is well modeled by this process. We have used our model to predict the time of the next CDM with accuracy exceeding both the baseline and a simpler predictor using the classical probabilistic approach.
The estimated probability is conservative when compared to the reality, and this might mean that the initial hypothesis of homogeneity might not be completely matched to the data. While the deviation is small, it is consistent, and it might indicate that in the last two days, there is a bias to predict a smaller number of CDMs. However, for the practical case of a high-stakes application like space awareness, it is always safer to be under-confident than over-confident.

\section*{Acknowledgments}
The authors of this paper would like to thank European Space Agency (ESA) for providing data, LARSyS-FCT Project UIDB/50009/2020 for the help, and to Neuraspace for supporting this research.

\bibliographystyle{named}
\bibliography{main}

\end{document}